\documentclass[conference]{IEEEtran}
\IEEEoverridecommandlockouts
\usepackage{cite}
\usepackage{amsmath,amssymb,amsfonts}
\usepackage{algorithmic}
\usepackage{graphicx, subcaption}
\usepackage{textcomp}
\usepackage{lipsum}
\usepackage{xcolor}
\usepackage{float}
\def\BibTeX{{\rm B\kern-.05em{\sc i\kern-.025em b}\kern-.08em
    T\kern-.1667em\lower.7ex\hbox{E}\kern-.125emX}}
\begin{document}


\title{Towards Precision Aquaculture: A High Performance, Cost-effective IoT approach}

\author{
\IEEEauthorblockN{Rafael R. Teixeira, Juliana B. Puccinelli\\ Luis Poersch, Marcelo R. Pias and Vin\'{\i}cius M. Oliveira}
\IEEEauthorblockA{\textit{FURG} - Rio Grande, RS, Brazil \\
{rafael, julianapuccinelli, mpias}@furg.br,\\
lpoersch@gmail.com, vinicius@ieee.org}

\and
\IEEEauthorblockN{Ahmed Janati, Maxime Paris}
\IEEEauthorblockA{\textit{BIOCEANOR}\\
Sophia Antipolis, France \\
ahmed.janati@bioceanor.com\\ maxime.paris@bioceanor.com}

}

\maketitle

\begin{abstract}

Demand for ocean-based high-quality and sustainable fish protein soared in the last decade. Unlike precision agriculture, aquaculture remains an under-equipped farming activity. To fully realise a vision of \textit{precision aquaculture}, smart IoT can offer the essential tools to local producers worldwide. This paper addresses the question of whether it is feasible to deploy a water quality sensor network in aquaculture for real-time user visualisation and decision making. The proposed system design and its implementation have been validated in a typical aquaculture farming facility. Preliminary results suggest the approach is feasible for short to medium LoRa communication range (up to 110m from outdoor water tanks). Although sensor data loss occurred, the end-to-end data path from the water sensors up to the user's monitoring app has been a positive outcome. The proposed IoT system approach provides the ground for a sustainable precision aquaculture~\footnote{This technology-based work has received funding from the European Union's Horizon 2020 research and innovation programme under grant agreement No 863034 (ASTRAL - All Atlantic Ocean Sustainable, Profitable and Resilient Aquaculture - https://www.astral-project.eu}. 

\end{abstract}


\section{Introduction}
Aquaculture is the farming of fish, crustaceans, molluscs, aquatic plants, algae, and other organisms. Activities in aquaculture have seen a global annual growth rate of 7\%, highlighting their increased relevance to animal protein production worldwide\footnote{FAO. 2018. The State of World Fisheries and Aquaculture 2018 - Meeting the sustainable development goals. Public report}. As a result, the North and South Atlantic production has received considerable government and policymakers attention from the '90s onwards. For instance, Brazil's government stimulus and private investments helped increase the  \textit{per capita} fish consumption in this country, thus leading to a net production of 707 thousand tons in 2015. Such tailored interventions have taken Brazil to 12th place in the world's aquaculture league. On a global scale, 
aquaculture fish production reached 82.1 million tons worldwide in 2018 according to an FAO report \cite{FAOReport}. The substantial social and economic impacts of aquaculture farming cannot be underestimated as a catalyst for new employment posts throughout the whole value chain. Such benefits reach the small producers located on both sides of the Atlantic,  including impoverished areas in low-to-middle income countries (e.g. communities in South America and Africa). Given the demand rise for high-quality animal protein to feed the world population, marine aquaculture is under tremendous pressure to accelerate production in a responsible, eco-friendly manner. A crucial enabler is the technology for real-time monitoring of aquatic organisms in an operational setting. Rigorous control of the environment's physical-chemical parameters is the first step towards an economically viable production. Unbalanced parameters can directly affect species' growth, generate pathogens and environmental conditions that may hinder animal reproduction and healthy growth. Aquaculture is an unpredictable biological system  in which adverse outcomes cannot be ignored in the management of a production system~\cite{valenti2021}. The European Commission adopted the term \textit{Aquaculture 4.0} to contemplate the transfer of new technologies investigated in other industries, including Internet of Things~(IoT), distributed computing, data analytics and artificial intelligence.  The Internet of Things (IoT) is a rapidly growing technological field, providing social and economic benefits for all geographical areas but particularly low-cost technology delivered to low-to-middle income countries. For example, traditional water monitoring in aquaculture follows a very laborious and time-consuming procedure whereby essential variables are measured. The high cost of the needed human resources severely limits the monitoring and control of the aquaculture facilities \cite{cloudaquaculture,prototypeofaquaculture}. 

IoT-driven monitoring and control in aquaculture is a challenging scenario in itself because of a number of constraints:
\begin{enumerate}
    \item Site	area:	anything	between	5	to	30	hectares;
    \item Difficult	access	(boats	or	off-road	vehicles);
    \item Power	supply:	none	in	open-water;	none	to limited	in	land-based facilities;
    \item Communication	coverage:	limited	to	none;
    \item Climate	and	adverse	weather	conditions;
    \item Lack	of	reliable and suitable monitoring technology.
\end{enumerate}

Aquaculture 4.0 offers tools that can be leveraged and applied to level up production systems' quality and effectiveness. Previous limitations can be overcome when reliable technologies monitor and identify any undesired condition or event of interest learnt from data. 
A key question to ask is the following one: 
\textit{how feasible are the design and deployment of a water quality sensor network that communicates data to an online platform for real-time visualisation and user decision making? ~\cite{anovelmethodology}}. 

This paper presents initial validation results of the \textit{AquaGreen} IoT system as an attempt to shed some light on the solution space for the question above. The specific contributions are:
\begin{itemize}
    \item System design specification and initial validation of cost-effective IoT monitoring system~(proof of concept) for real-world aquaculture farming. 
    \item In-loco validation of real-time, low-power long-range data communication based on LoRaWAN.
    
\end{itemize}

This paper takes further Pister~\cite{Pister} \textit{smartdust} vision to develop a bottom-up approach for sensor application development. It departs, however, from the biased view that sensor technology can be easily deployed into the target environment (e.g. self-organising sensors "thrown" from airplanes \cite{Pister}). Unlike this, the practicalities of placing IoT technology in the real world should be fully appreciated for the \text{smartdust} vision to realise. 
The \textit{AquaGreen} described in this paper contributes to sustainable aquaculture in tackling a set of specific U.N.'s Sustainable Development Goals\footnote{United Nations Sustainable Development Goals: https://sdgs.un.org/goals}: SDG 2 - Zero Hunger, SDG 9 - Industry, Innovation, Infrastructure, SDG 11 - Sustainable and Connected Communities around the Globe and SDG 14 - Life below Water.

\subsection{Water Quality Monitoring}
High-quality water is a crucial requirement across the board for marine aquaculture, where production performance is often excelled in those cases of tight monitoring and control.  Poor management of basic physical-chemical parameters (temperature, dissolved oxygen, pH, turbidity and ammonia) can result in significant losses leading to a production reduction of anything between 20\% and 40\% \cite{iotaquaculture}. A significant drop in dissolved oxygen levels, mainly considering an intensive production, can lead to an undesirable loss in shrimp farms. A few catastrophic animal losses have been experienced at FURG's Marine Aquaculture Facility (FURG-EMA), in some cases the slow user reaction (more than 30 minutes) affected almost the entire cycle of shrimp production. Temperature variation is one of the most critical factors to keep a tight monitoring-control scheme. All the species' physiological and behaviour related activities, such as breathing, digestion, and feeding, are directly linked to the water temperature levels. The higher the temperature, the greater the level of animal activities. As a result, oxygen consumption is increased likewise. 
Monitoring essential water quality variables is a massive step towards aquaculture enhanced production. The term performance here takes a broader view that puts side-by-side production and eco-friendly sustainable processes.  

\subsection{Aquaculture IoT Kits: User Requirements}
The vision of IoT as advocated in~\cite{xia2012internet} comprising ubiquitous embedded A.I. sensors and cooperating objects~\cite{Ollero} lead to a highly distributed network of smart sensor devices communicating with both end-users and peer devices. Such a vision has been presented to end-users in a workshop held in early 2021 at FURG-EMA (Brazil). Further requirements were also solicited from the H2020 ASTRAL project consortium. Despite the users' needs for new types of advanced monitoring, say microplastic contaminants and animal biomass, the list that follows capture design issues of IoT applied to aquaculture applications:
\begin{itemize}
\item Essential water monitoring variables: temperature, dissolved oxygen, pH, turbidity and ammonia; 
\item Measurement sampling rate: one every second; 
\item Real-time data communication: measurement data to reach the end-user with due time for control intervention in case of detected anomalies; 
\item Scalability: solution to function in sparse and remote located land-based aquaculture farms; 
\item Limited power availability: a power-efficient system for long-lasting data gathering and communication.  
\end{itemize}

This list calls for a system capable of monitoring essential variables even when farmers are distant from the breeding and production sites. In addition, the system enables intelligent real-time control and predictive anomaly detection that maximises the producer awareness. A.I models can enhance	the	knowledge	of	local	producers,	lower	existing	barriers	of	aquaculture	activities	by	the	society at large.

\subsection{Recirculation Aquaculture System - RAS}
Aquaculture is still an under-equipped sector in terms of intelligent systems. However, regardless of size and scale, water quality remains the common ground in farming facilities worldwide. RAS (Recirculation Aquaculture System) is a method in which the production system is continuously treated and reused, allowing the total or partial reuse of water, thus decreasing water waste in a production cycle. In particular, the RAS approach develops a treatment scheme system of the effluents generated in the cultivation through induced water flow in the land-based tanks. In contrast, offshore marine aquaculture, farms located at a distance from the coast, have experienced rapid growth due to limited land resources and high competition for their use in some economic sectors. In addition, offshore facilities employ specialised labour to achieve robustness and profitability. Open sea farming offers sustainability more inherently as strong ocean currents provide circulation and handling of waste. However, long-term restrictions to offshore farming have made land-based RAS popular in the last decade or so. Less exposure to diseases and environmental threats are regarded as crucial benefits. A few effective industrial processes, tailored training and sensor technology, are expected to bring professionalism to producers. In addition, the use of water recirculation in closed land-based systems (RAS) should become affordable for widespread sector penetration. 

\section{Related Work}
Lafont (et al.) \cite{iotaquaculture} argues that IoT applied to aquaculture is embryonic still. This application has the potential to create a coherent development front towards the concept of \textit{precision aquaculture}. In that sense, the IoT contribution goes beyond the networked wireless nodes. Sensor data analytics play a key role in handling complex data-oriented challenges, including anomaly detection, predictive facility operation and maintenance - to cite a few.  The authors in~\cite{aquaculture4-0} discuss the easy-to-use aspects of IoT tools when compared to the cumbersome tools that aquaculture producers commonly use in day-to-day practices. The same work observes that such tools have to be intelligent, accessible, easy to implement, reliable and highly efficient. Both studies rely on cost-effective wireless sensors to put together an infrastructure where easy-to-use UIs graphically present to end-users changes in monitored water pH, dissolved oxygen, and other variables. 

This paper goes beyond when it addresses the practical and subtle challenge of a real-world pilot deployment. Specifically, this paper contributes with a better understanding of the issues associated with sensor data communication quality, communication range, and data volume in a small yet powerful real-world deployment in a aquaculture facility. Possible control and actuation in the target environment allow for a diverse set of smart interventions such as adding stabilisation chemicals to the water.

\section{System Approach}
IoT enhanced with artificial intelligence has potential for widespread use in farming activities, particularly aquaculture. Nevertheless, researchers and practitioners should overcome problems highly dependent on taking technology to the target environment (e.g. limited power and communication coverage in both land-based and offshore facilities). In conjunction with the recirculation production system (RAS), sensors, and actuators, machine learning will pave the way to the vision of smart precision aquaculture. FURG is active in aquaculture research for new systems, methodologies and species. FURG's Marine Aquaculture Station (EMA) is a complete production facility catering for the species: Whiteleg shrimp~(Litopenaeus vannamei), Nile tilapia~(Oreochromis niloticus), Crassostrea tulipa~(Crassostrea gasar), Sea lettuce (Ulva S.P.) and Salicornia neei. The 2,800 sqm facility offers an integrated multi-trophic system~(IMTA) where two or more aquatic species from different food chain levels are farmed together so that waste from one species can be used as a resource to another, thus increasing circularity and reducing waste. However, the facility provides limited power and cellular communication for uplink internet connectivity. This scenario is representative of a traditional land-based marine aquaculture fishery facility. This work focuses on the features and limitations of a  networked sensor system, its implementation and preliminary prototype validation in the field. 

\begin{figure}[htb]
    \centering
    \resizebox{\linewidth}{!}{
        \includegraphics{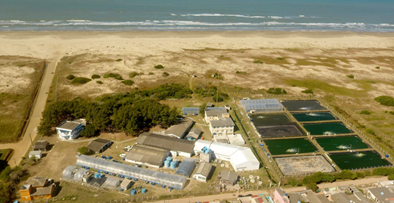}}
    \caption{Marine Aquaculture Station - FURG EMA}
    \label{fig:ema}
\end{figure}

\begin{figure}[h!]
    \centering
    \resizebox{\linewidth}{!}{
    \includegraphics[scale=0.35]{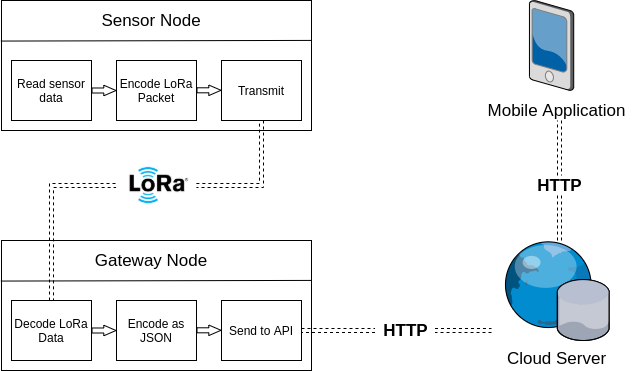}
    }
    \caption{Simplified Network Architecture}
    \label{fig:network-architecture}
\end{figure}

The first architecture version envisaged many unnecessary component-level functionalities such as several data sources, edge data computing and in-node versus cloud processing aspects. The second version, now simplified after consultation with end-users, comprises two levels of data interaction, namely local and external. The local interaction consists of sensor-actuator nodes structured in a local communication layer and a gateway node capable of gathering sensor data to be relayed to an external entity. In contrast, sensor data is stored and streams processed in the external layer (e.g. servers in the cloud). Authorised users access real-time analysed data for decision-making, such as increasing water aeration in a specific tank). Figure \ref{fig:network-architecture} illustrates the proposed network architecture.

\subsection{Local Layer}
This layer addresses the communication from sensor nodes to the gateway over a Low Power Wide Area Network~(LPWAN). LoRa has been because of its small energy consumption footprint and enhanced communication coverage ~\cite{SINHA201714}. Nodes sample specialised water quality sensors. Collected sensor data are encoded in a LoRa packet and transmitted to an in-range gateway node. Received sensor data (encoded in a JSON template), are forwarded via HTTP to an external cloud server.

\subsection{External Layer}
Gateway nodes receive external connectivity either through an internet-connected WiFi router or cellular network, whichever is present at the aquaculture facility. In a few remote sites where there is limited to none internet access, it is anticipated that the gateway node enters into a delay tolerant network mode~(DTN~\cite{kevin2003}). This particular DTN setup requires end-user intervention to decide how to forward the stored data through some type of device-to-device transfer or data off-loading via a smartphone cellular interface.  In this work, gateway nodes communicate sensor data to external domain servers located in a commercial cloud infrastructure (e.g. AWS EC2 instances). A commercial ISP provides internet connectivity to the gateway node.  The external cloud server offers a customised sensor data API that receives JSON formatted data and stores them into a NoSQL database for improved scalability in volume handling. The server API also provides endpoints for clients to consume the stored sensor data using a customised scheme with a regular expression scheme for data searching. 

\subsection{Embedded Hardware and Software}
System extensibility is a prime design choice for sensor-enabled aquaculture applications. The intention here is to efficiently extend the IoT Kit functionality by supporting multiple communication protocols, analogue/digital sensors and actuators, and onboard co-processing units. The embedded software is the one that should do most of the work to load up software modules for new types of integration. Also, system power management plays a role in optimising battery charge. This led the team to choose an ESP32 board comprised of two 32-bit CPU cores and an ultra low power co-processor. ESP32 supports a deep sleep mode that brings down the current draw to a level under 10µA. The SX1276 LoRa transceiver module creates a powerful sensor node capable of multi-communication that gives flexibility for comparison and validation in a variety of aquaculture farming facilities (e.g. from resourceful to limited power and communication support).  Also, the chosen node supports WiFi, BLE and ESP-Now considered useful complementary communication technologies in challenging IoT contexts. The idea is to provide functionality for edge processing. For instance,  anomaly detection algorithms for water quality data can be embedded into the ESP32. The recently developed TinyML framework based on Google's TensorFlow Lite for Microcontrollers\footnote{https://www.tensorflow.org/lite/microcontrollers} offers a new dimension for embedded machine learning in the proposed sensor node.  

\subsection{Experimental Setup: Farm Radio Surveying}
The experiments below were carried out to survey a typical aquaculture farm (e.g. FURG-EMA):
\begin{itemize}
\item \textbf{Experiment 1:} to validate the maximum range for LoRa communication within the vicinity of the facility. The end-users carefully chose nine location points  used for the LoRa transmitter side. Sensor data packets were sent from each of the locations towards a fixed positioned LoRa receiver (base station). Average received signal strength (RSSI) was recorded. 
\item \textbf{Experiment 2:} to provide a first-hand evaluation of the end-to-end application data path (i.e. in-water sensing to a mobile app). The setup included integrating a calibrated water temperature sensor attached to a sensor node, a gateway node, a cloud server API and a mobile app. The latter consumes time-series water temperature data to be overlaid in a customised user monitoring app. 
\end{itemize}

\section{Initial Results}

\subsection{LoRa Range Issues (Experiment 1)}

LoRaWAN is a widely adopted solution in IoT networks due to its capability to connect to a wide area with many devices and still offer a low power communication \cite{attia}. In this setup, LoRa is used as the physical layer protocol of a LoRaWAN network. However, the low data transmission rates present a challenge for IoT applications, especially those relying on image-based sensors. In its low-power basic setup, Aquaculture IoT should rely on time-series sensor data that is tiny enough to be transported through a LoRa-based network. This experiment aims at determining whether LoRa communication can provide sufficient network coverage in the field. A pair of TTGO LoRa32 v1.0 development boards were used to implement the sensor nodes. Each board includes an ESP32 MCU, an SX1276 LoRa Transceiver, a 3dBi antenna and a 0.96inch OLED display. The system has a JST battery plug and a battery charging unit. A Li-ion 3.7V battery powers up the sensor nodes. 

Sensor nodes were programmed interchangeably as LoRa transmitters and receivers. The chosen frequency of 915MHz complies with the local radio regulations even though radio interference is unlikely typical deployment sites. The system operated using SF7 spreading factor at a transmission power of 17dB. Other system configurations could undoubtedly be used, but this one provided the base for a field evaluation. The receiver node powered from a reliable power source was placed at a distance from the fishery tank. The transmitter nodes had their positions varied one at a time over the set of nine possible locations.  The transmitter sent a batch of 10 LoRa packets to the receiver node. Received Signal Strength Indication (RSSI) values associated with received packets were calculated based on the following communication power loss model \cite{loraDocumentation}: 
\begin{equation*}
    L_{fs} = 32.45 + 20log(D) + 20log(f)
\end{equation*}
where $L{fs}$ is the free space loss in decibels, $D$ is the distance (in km) between transmitter and receiver nodes, and $f$ is the frequency in MHz.

\begin{figure}[hbt]
    \centering
    \resizebox{0.5\columnwidth}{!}{
    \includegraphics{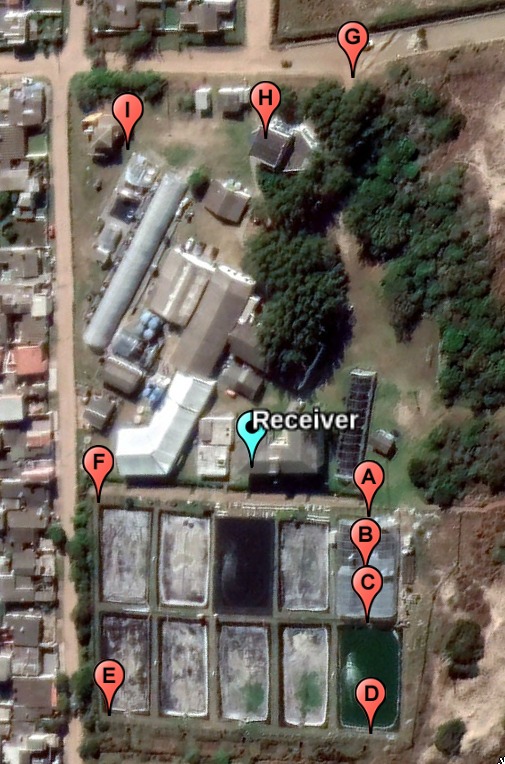}
    }
    \caption{Satellite view of measurement points}
    \label{fig:map2}
\end{figure}

Figure~\ref{fig:map2} presents the testing locations explored at the aquaculture farm~(FURG-EMA). The locations were carefully selected to provide a more realistic survey of the communication range and possible blind spots for sensor deployment. Laser-based surveying and satellite images were also used to calculate approximate linear distances between transmitter and receiver nodes. 

Points $A$, $B$ and $C$ are located under a greenhouse roof structure covering some of the facility RAS water tanks (Figure \ref{fig:map2}). Points $D$, $E$ and $F$, located in the southern limits of the perimeter, cover the farm's external tanks  
Additionally, northern location points  ($I$, $H$ and $G$) provided extended coverage for a complete site surveying. Results obtained from these experiments (Table \ref{table:1}) suggest that the selected LoRa module can cover the communication between points located at the RAS greenhouse tanks and external fishery water tanks. However, this initial configuration was unable to cover the entire site area; for instance, the testing points $G$, $H$ and $I$ could not be reached from the LoRa transmitter. A possible cause for this issue is the actual distance of over 110m, several buildings  in the middle (point $I$) and a small tropical forest (points $G$ and $H$) as shown in Figure~\ref{fig:map2}). Experimenting with different antenna design and radio frequencies could increase the transmitter's range. IoT-based site surveying will not be an easy task in real-world aquaculture farm deployments. Satellite-based images can assist the survey task though. The issues that arise from the farm's remote location, either in offshore or in-land facilities, are far more challenging to address as indicated in (Table~\ref{table:1}). Similarly to agricultural rural areas, aquaculture farming requires a systematic approach for site radio surveying that relies on fresh satellite images to guide a time-consuming yet needed manual intervention. 
\begin{table}[hbt]
\centering
\caption{\\LoRa Range Test Results}
\label{table:1}
\scalebox{1.25}{%
\begin{tabular}{ccc}
\hline
Location & Linear Distance & Average RSSI \\ \hline
A        & 43m             & -108dBm      \\ 
B        & 55m             & -110dBm      \\
C        & 70m             & -109dBm      \\
D        & 104m            & -109dBm      \\
E        & 102m            & -110dBm      \\
F        & 56m             & -107dBm      \\
G        & 143m            & no signal    \\
H        & 117m            & no signal    \\
I        & 120m            & no signal    \\ \hline \\
\end{tabular}%
}
\end{table}

\subsection{End-to-End Architecture Implementation (Experiment 2)}

Putting aside the deployment challenges, the results obtained in Experiment 1 encouraged the research team and end-users to continue working towards a preliminary implementation of a functional system. To fully implement and test the proposed system architecture, the sensor node design included capabilities for the adverse weather conditions of an open area environment. The functional set of sensor nodes (Kits), embedded gateway, cloud server software and mobile app comprise the end-to-end IoT Kit. This has been termed the \textit{AquaGreen system}. Local access restrictions to the aquaculture facility imposed in February 2021 due to the worsening of the COVID-19 pandemic in Brazil limited the planned experiment scope. Although the system network layer (Experiment 1) has been assessed in normal operating conditions in December 2020, the AquaGreen system had no other option but to be tested in an outside swimming pool. This lab setup still offered rigorous conditions to validate technical design assumptions. 

\begin{figure}[htb]
    \centering
    \begin{subfigure}[b]{0.35\linewidth}        
        \centering
        \includegraphics[width=\linewidth]{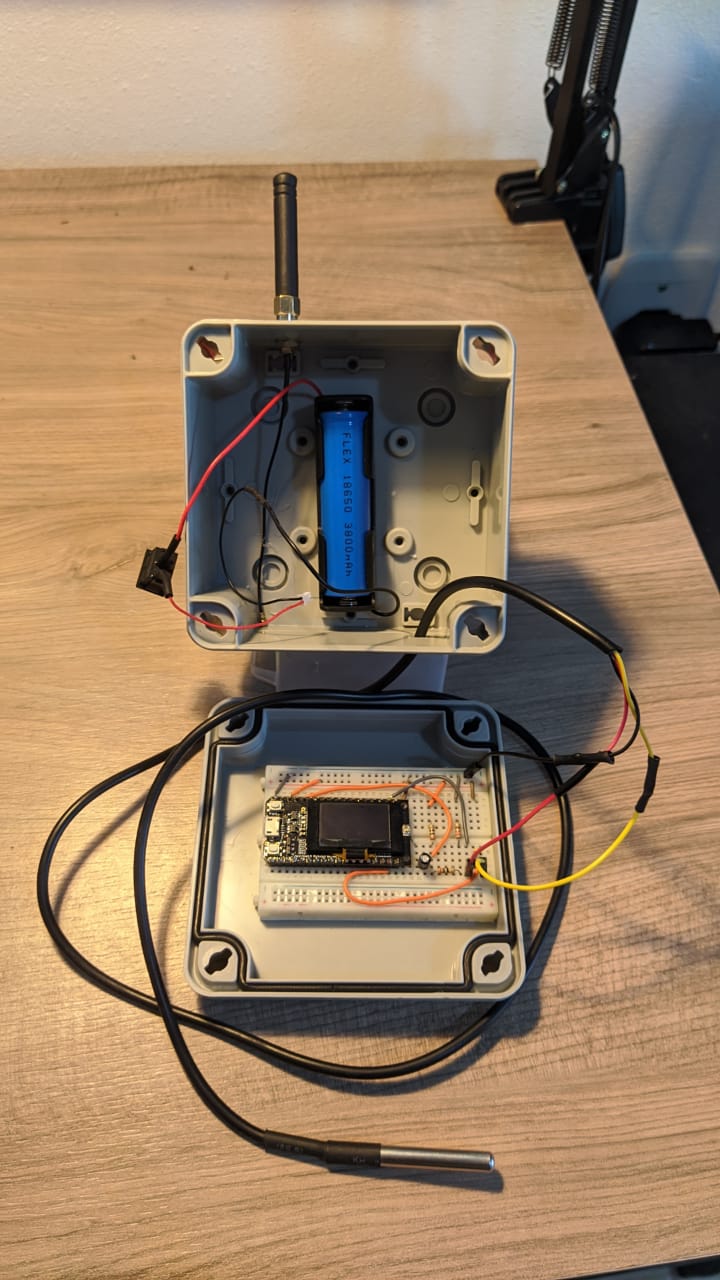}
        \caption{Internal view of the sensor node}
        \label{fig:sensor-node-1}
    \end{subfigure}
    \begin{subfigure}[b]{0.35\linewidth}        
        \centering
        \includegraphics[width=\linewidth]{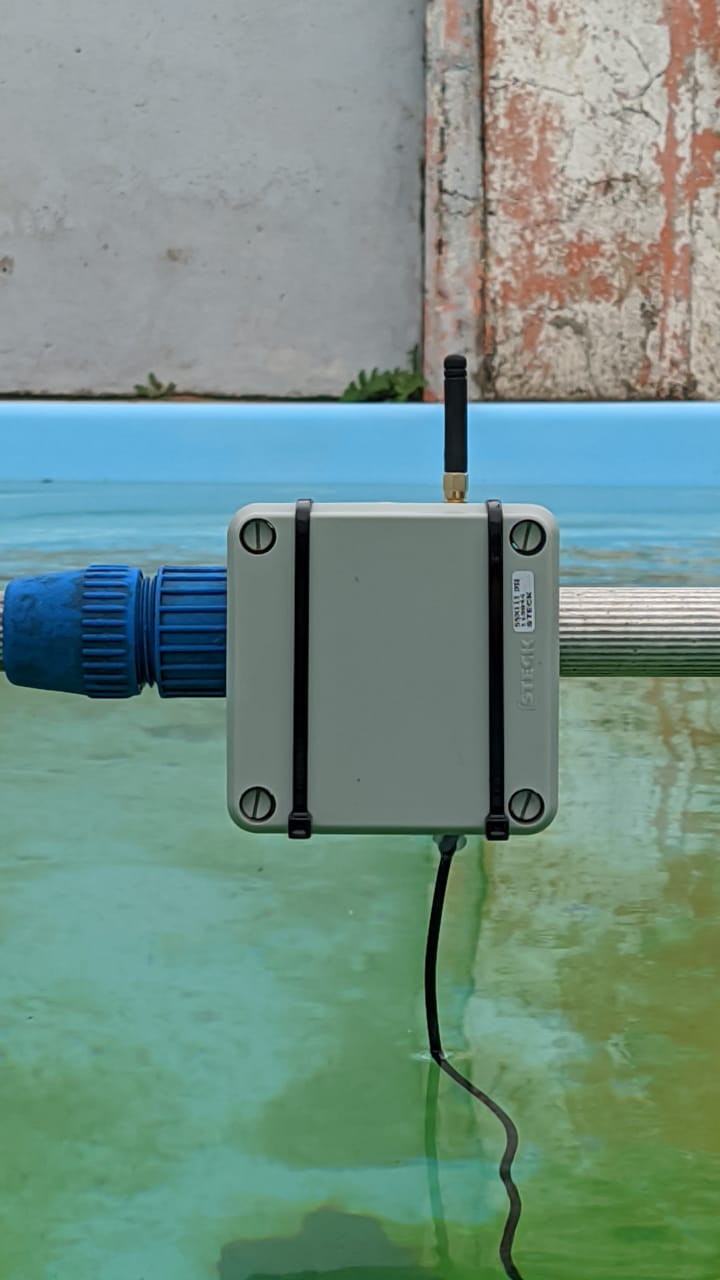}
        \caption{Sensor node encasing during the test}
        \label{fig:sensor-node-2}
    \end{subfigure}
    \caption{AquaGreen Sensor node prototype}
    \label{fig:sensor-node}
\end{figure}

A DS18b20 temperature sensor continuously monitors the pool's water temperature for a specific time window. Waterproof casing enclosed a few system components, including the ESP32 sensor node board, 3.7 Li-ion battery and cables used for the external radio antenna and attached sensor probe. The development board had been attached to the internal wall through a strong double-sided adhesive tape. Figure \ref{fig:sensor-node} shows a picture of the \textit{AquaGreen} system.

To obtain the maximum battery life, the sensor node is kept in deep sleep mode most of the time. During this period, the temperature sensor is also powered down. On deep sleep mode, the overall system power consumption is reduced to around 10µA. The chosen system duty cycle contemplates that every 10 minutes, the sensor node wakes up from deep sleep to sample the sensors. The sample data is encoded in a LoRa packet and sent to the gateway node. The LoRa packet holds four key-value pairs of data samples (i.e. real-clock timestamp, sensor node id, instantaneous water temperature sensor sample and battery voltage level). Due to local regulations, only a maximum of 1\% duty cycle is permitted in LoRa communication. The system can send a packet (30 bytes payload plus 15 bytes header) every 10 seconds. To comply with the local rules, the actual packet rate is one every 10 minutes. As the final setup step, the sensor node was placed above the water in the swimming pool but having the temperate sensor probe under the water in a 1.2 water depth approximately (Figure~\ref{fig:sensor-node-2}).


The gateway node has been implemented with the same hardware used in range in Experiment 1. The key difference lies on the embedded software, which now implements the system architecture functional blocks. Specifically, in this case, deep sleep mode in the gateway node was unnecessary due to the availability of USB power throughout the experiments performed. 

The gateway node acts as a LoRa receiver. Whenever a LoRa packet arrives from the sensor node, the gateway unpacks it and encodes the data in a JSON file. The node uses this document-oriented data format and an authentication token to create a payload for an HTTP POST request to the cloud server. The gateway node received Internet connectivity through the ESP32's WiFi interface. The real-time monitoring mobile app allows the user to authenticate into the system, providing thus authorised access to the most recent sensor readings. The cloud server functions as a data forwarding path to the end-users. It securely stores the data received from the gateway nodes into a document-oriented database. Then, the data is provided to a mobile app. Industry-grade security has been used to secure the end-to-end data path with carefully selected cloud-based services. The current server implementation uses Javascript, node.js and MongoDB database. The server also handles user authentication through JSON Web Tokens (JWT) to guarantee that only authenticated entities are authorised to send and receive data from the cloud server. Overall, the results of this simplified yet necessary experiment have been positive. The \textit{AquaGreen} system ran for a non-stop period of 66 hours up to the point of fully discharging batteries in the sensor nodes. A total of 392 successful sensor readings were stored on the database during this time interval. Unfortunately, a packet loss of 2.3\% has been observed.  Although this has not impacted the application, it calls for further investigation into whether a reliable communication protocol should be used. 


\section{Conclusion}

This work presented preliminary results of a real-world pilot in a challenging environment for sensor and actuator networks. In addition, design principles and best practices for IoT aquaculture application development were presented. Experimental results shed some light on the feasibility of a water quality sensor network that communicates data to an online platform for real-time visualisation and user decision making. The first-hand results showed the challenges for low-power long-range communication in a typical aquaculture farm environment where buildings and tropical forests surround external tanks. There are still many improvements to be made to the overall system. Experimental with different radio and antenna configurations have the potential to extend the communication range. A solar-based power source can offer a viable solution to water tanks with no energy supply. As for the decision-making software, analytical insights and predictions based on previous time-series sensor recordings are under development. Predictive monitoring and alerts to the users solve a significant issue: prompt user reaction when water quality deteriorates. It only takes around 30 minutes for a complete production loss in RAS-based tanks. Real-time processing is crucial in this case. Although the training of neural networks remains a cloud-based GPU task, TinyML and Tensorflow Lite for Microcontrollers provide a fast inference engine that can be rightly embedded into the proposed system. 

\bibliographystyle{IEEEtran}
\bibliography{bibliografia.bib}

\begin{thebibliography}{10}
\providecommand{\url}[1]{#1}
\csname url@samestyle\endcsname
\providecommand{\newblock}{\relax}
\providecommand{\bibinfo}[2]{#2}
\providecommand{\BIBentrySTDinterwordspacing}{\spaceskip=0pt\relax}
\providecommand{\BIBentryALTinterwordstretchfactor}{4}
\providecommand{\BIBentryALTinterwordspacing}{\spaceskip=\fontdimen2\font plus
\BIBentryALTinterwordstretchfactor\fontdimen3\font minus
  \fontdimen4\font\relax}
\providecommand{\BIBforeignlanguage}[2]{{%
\expandafter\ifx\csname l@#1\endcsname\relax
\typeout{** WARNING: IEEEtran.bst: No hyphenation pattern has been}%
\typeout{** loaded for the language `#1'. Using the pattern for}%
\typeout{** the default language instead.}%
\else
\language=\csname l@#1\endcsname
\fi
#2}}
\providecommand{\BIBdecl}{\relax}
\BIBdecl

\bibitem{FAOReport}
\BIBentryALTinterwordspacing
Why iot, big data \& smart farming are the future of agriculture. [Online].
  Available:
  \url{http://www.fao.org/e-agriculture/news/why-iot-big-data-smart-farming-are-future-agriculture}
\BIBentrySTDinterwordspacing

\bibitem{valenti2021}
W.~C. Valenti, H.~P. Barros, P.~Moraes-Valenti, G.~W. Bueno, and R.~O. Cavalli,
  ``Aquaculture in brazil: past, present and future,'' \emph{Aquaculture
  Reports}, vol.~19, p. 100611, 2021.

\bibitem{cloudaquaculture}
M.~{Cordova-Rozas}, J.~{Aucapuri-Lecarnaque}, and P.~{Shiguihara-Juárez}, ``A
  cloud monitoring system for aquaculture using iot,'' in \emph{2019 IEEE
  Sciences and Humanities International Research Conference (SHIRCON)}, 2019,
  pp. 1--4.

\bibitem{prototypeofaquaculture}
K.~S.~S. {Javvaji} and M.~A. {Hussain}, ``Prototype of aquaculture using iot
  technologies,'' in \emph{2020 11th International Conference on Computing,
  Communication and Networking Technologies (ICCCNT)}, 2020, pp. 1--4.

\bibitem{anovelmethodology}
T.~{Abinaya}, J.~{Ishwarya}, and M.~{Maheswari}, ``A novel methodology for
  monitoring and controlling of water quality in aquaculture using internet of
  things (iot),'' in \emph{2019 International Conference on Computer
  Communication and Informatics (ICCCI)}, 2019, pp. 1--4.

\bibitem{Pister}
J.~M. Kahn, R.~H. Katz, and K.~S.~J. Pister, ``Emerging challenges: Mobile
  networking for “smart dust”,'' \emph{Journal of Communications and
  Networks}, vol.~2, no.~3, pp. 188--196, 2000.

\bibitem{iotaquaculture}
M.~{Lafont}, S.~{Dupont}, P.~{Cousin}, A.~{Vallauri}, and C.~{Dupont}, ``Back
  to the future: Iot to improve aquaculture : Real-time monitoring and
  algorithmic prediction of water parameters for aquaculture needs,'' in
  \emph{2019 Global IoT Summit (GIoTS)}, 2019, pp. 1--6.

\bibitem{xia2012internet}
F.~Xia, L.~T. Yang, L.~Wang, and A.~Vinel, ``Internet of things,''
  \emph{International journal of communication systems}, vol.~25, no.~9, p.
  1101, 2012.

\bibitem{Ollero}
M.~B. P. J. M. A. O.~A. Wolisz, \emph{Cooperating Embedded Systems and Wireless
  Sensor Networks}.\hskip 1em plus 0.5em minus 0.4em\relax Willey, 2008.

\bibitem{aquaculture4-0}
C.~{Dupont}, P.~{Cousin}, and S.~{Dupont}, ``Iot for aquaculture 4.0 smart and
  easy-to-deploy real-time water monitoring with iot,'' in \emph{2018 Global
  Internet of Things Summit (GIoTS)}, 2018, pp. 1--5.

\bibitem{SINHA201714}
R.~S. Sinha, Y.~Wei, and S.-H. Hwang, ``A survey on lpwa technology: Lora and
  nb-iot,'' \emph{ICT Express}, vol.~3, no.~1, pp. 14--21, 2017.

\bibitem{kevin2003}
K.~Fall, ``A delay-tolerant network architecture for challenged internets,'' in
  \emph{Proceedings of the 2003 Conference on Applications, Technologies,
  Architectures, and Protocols for Computer Communications}, ser. SIGCOMM
  '03.\hskip 1em plus 0.5em minus 0.4em\relax New York, NY, USA: Association
  for Computing Machinery, 2003, p. 27–34.

\bibitem{attia}
T.~Attia, M.~Heusse, B.~Tourancheau, and A.~Duda, ``Experimental
  characterization of lorawan link quality,'' in \emph{2019 IEEE Global
  Communications Conference (GLOBECOM)}, 2019, pp. 1--6.

\bibitem{loraDocumentation}
\BIBentryALTinterwordspacing
Lora documentation - free space losses. [Online]. Available:
  \url{https://lora.readthedocs.io/en/latest/\#free-space-losses}
\BIBentrySTDinterwordspacing

\end{thebibliography}
\end{document}